\documentclass{JHEP}

\def\la{\lambda}

\def\a{\alpha}
\def\b{\beta}

\def\d{\delta}

\def\g{\gamma}

\def\o{\omega}

\def\O{\Omega}

\def\sg{\sqrt{-g}}

\newcommand{\eq}{\begin{equation}}
\newcommand{\feq}{\end{equation}}
\newcommand{\eqn}{\begin{eqnarray}}
\newcommand{\feqn}{\end{eqnarray}}
\newcommand{\arr}{\begin{eqnarray*}}
\newcommand{\farr}{\end{eqnarray*}}
\newcommand{\beq}{\begin{equation}}
\newcommand{\eeq}{\end{equation}}
\newcommand{\bea}{\begin{eqnarray}}
\newcommand{\eea}{\end{eqnarray}}

\def\sg{\sqrt{-g}}

\def\d{\delta}
\def\e{\Phi}

\def\lb{\label}

\title{2D Black Holes, Conformal Vacua and CFTs on the Cylinder}

\author{Mariano Cadoni and  Paolo Carta \\
Universit\`a degli Studi di Cagliari, Dipartimento di
Fisica and\\ INFN, Sezione di Cagliari, Cittadella Universitaria 09042,
Monserrato, Italy. \\
Email: \email{mariano.cadoni@ca.infn.it}, \email{paolo.carta@ca.infn.it}}

\abstract{We investigate two-dimensional dilaton gravity models with a
power-law dilaton potential, whose black hole solutions contain, among
others, the dimensional reduction of the Schwarzschild black hole, the
Anti-de Sitter black hole and Rindler spacetime. We show that the
ground state of these models satisfies simple transformation laws
under the $SL(2,R)$ conformal group. We use these transformation laws
to explain the scaling behavior of the thermodynamical parameters of
the black hole and the nonextensivity of the thermodynamical system.
The black hole thermodynamical behavior, in particular its entropy, is
reproduced by a mapping into a two-dimensional conformal field theory
on the cylinder.}

\preprint{INFNCA-TH0106}

\begin{document}

\section{Introduction}
Recent developments have pointed out the relevance of nonextensivity 
in understanding the microscopic origin of the entropy in 
gravitational systems \cite{verlinde,cosmo,black}.
In particular the Cardy-Verlinde formula \cite{verlinde} gives an expression 
of the entropy in terms of the energy and of the (sub-extensive) Casimir 
energy of the system, which is valid for all dimensions. 
Originally proposed in a cosmological context, the Cardy-Verlinde 
formula has been readily extended to other gravitational systems, in 
particular black holes \cite{black}. It has been successfully used for 
Anti-de Sitter (AdS)/ Schwarzschild black holes , CFTs with AdS
duals, charged black holes and  asymptotically flat 
black holes in any dimension.

For two-dimensional (2D) gravitational systems (more in general systems 
that admit 2D CFTs as duals) one can make use, directly, of the Cardy 
formula \cite{cardy} that gives the entropy of  a CFT in terms of the 
central charge $c$ and the eigenvalue of the Virasoro operator $l_{0}$.
However, this is possible only for 2D  systems for which 
one can explicitly show (e.g using the AdS/CFT 
correspondence) that they are 
in correspondence with a 2D CFT \cite{CM99,strominger}.
Even in this most favorable  case the use of 
the Cardy formula for the computation of the entropy of the 
gravitational system is far from trivial. $c$ and $l_{0}$ have to be 
expressed in terms of the gravitational parameters, an 
operation that sometimes turns out to  be very hard \cite{CCCM}.
The problem becomes much more intricate when a duality of the 
gravitational model with a CFT cannot be shown explicitly to hold (e.g. 
asymptotically flat 2D black holes).  

In this paper we consider two-dimensional dilaton gravity models with
a power-law dilaton potential. This class of models is particularly
interesting because it shows up in several contexts: dimensional
reduction to $D=2$ of the D-dimensional Schwarzschild solution (see for
instance \cite{CA}), effective theories for dilatonic 0-branes
\cite{youm,CCCM}, AdS$_{2}$ black hole \cite{CM95,CM99},
Callan-Giddings-Harvey-Strominger black hole (CGHS) (or its Weyl
rescaled version, the Rindler spacetime) \cite{CGHS,CM95a}.  We show
that the ground state of the model, although in general not invariant,
satisfies simple transformation laws under the action of the $SL(2,R)$
conformal group (Sect. 2). These transformation properties 
are used to explain the scaling behavior of the
thermodynamical parameters and the violation of the Euler identity,
which codify the non-extensivity of the thermodynamical system (Sect.
3).  The black hole is put in correspondence with a 2D CFT on the
cylinder.  The black hole mass is identified with the Casimir energy
of the cylinder vacuum and the Hawking temperature is given in terms
of the length of the compact dimension of the cylinder. The entropy of
the 2D black hole is reproduced using the Cardy formula (Sect. 4).  As
particular case we consider the Rindler spacetime. We point out that
our method can be used to give mass and entropy to the Rindler
spacetime in terms of the Casimir energy and entropy of a 2D CFT
(Sect. 5).

\section{Conformal properties of the ground state}
A particular interesting class of 2D dilaton gravity
models is represented by the action
\begin{equation}\label{action}
A={1\over2}\int \sg \, d^2x\, \left(\e R+\la^{2} V(\e)\right),
\end{equation}
with a power-law dilaton potential
\begin{equation}\label{pot}
V(\e)=(a+1)\e^{a},  \qquad a>-1\nonumber.
\end{equation}
The action (\ref{action}) emerges as 2D effective action both for 
D-dimensional Einstein gravity and dilatonic 0-branes in the string 
theory
context. D-dimensional, spherically symmetric, gravity reduces, after 
dimensional reduction of the (D-2)-dimensional sphere of radius 
$r=\e^{-a}$ to the model (\ref{action}) with $a=1/(2-D)$ \cite{CA}.
In particular for $a=-1/2$ our model describes the four-dimensional
Schwarzschild solution of general relativity. 
In the dual-frame, 
dilatonic 0-branes solutions of D-dimensional supergravity  
admit a (near-horizon) effective
description in terms of a two-dimensional dilaton gravity model
with a power-law potential \cite{youm,CCCM}.
The models  under consideration have also an intrinsic interest as 2D 
theories of gravity. For $a=1$ Eq. (\ref{action}) becomes the 
Jackiw-Teitelboim  model, which plays a crucial role in the 
Anti-de Sitter/Conformal Field Theory (AdS/CFT)  correspondence 
in two dimensions \cite{CM99}.  For $a=0$ we get the Weyl-rescaled 
version of the CGHS model \cite{CGHS,CM95a}, 
which 
admits the  Rindler spacetime as non-trivial dynamical solution
\cite{CM95a}.

In the Schwarzschild gauge, the general black hole solution of the 
model (\ref{action}) is
\bea\lb{bh}
ds^2 &=& -\left(\e^{a+1}-{2M\over\la}\right)dt^2 + 
\left(\e^{a+1}-{2M\over\la}\right)^{-1}dr^2\,,\nonumber\\
\Phi&=&\la r\,,
\eea
where $M$ is the Arnowitt-Deser-Misner  mass of the solution.
For $a\neq 0,1$ the solution has a curvature singularity shielded by 
an  event horizon at $\e=\e_{h}= \left({2M\over \la}\right)^{1/(a+1)}$.
For $a=1$ the solution describes 2D AdS black holes \cite{CM95}, whereas 
for $a=0$  Rindler spacetime endowed with a nontrivial dilaton 
\cite{CM95a}.

Let us now investigate the conformal symmetries of the $M=0$ black hole 
vacuum solution. For generic values of the parameter $a$ the $M=0$ 
solution is not maximally symmetric, it admits only one Killing vector 
generating time-translations. Only for $a=0,1$ when the   spacetime 
becomes Minkowskian, respectively,  AdS$_{2}$, the isometry group is 
maximal and, in the case of AdS$_{2}$, is the conformal group 
$SO(1,2)\sim SL(2,R)$.
Nonetheless, we expect  simple 
transformation laws of the vacuum solution under the action of the 
conformal group $SL(2,R)$, owing to its  power-law dependence on the 
coordinate $r$.

We will first consider $a\neq 0$, the case $a=0$ will be considered 
separately.
In the conformal gauge and using light-cone coordinates the $M=0$ 
solution in Eq. (\ref{bh}) takes the form,

\bea\lb{bhcg}
ds^2 &=& 2g_{+-}dx^{+}dx^{-}=\nonumber\\
&=&- \left[{|a|\la\over 2}(x^{+}-x^{-})\right]
^{-(1+{1\over a})}dx^{+}dx^{-}\nonumber\\
\e&=&\left[{|a|\la\over 2}(x^{+}-x^{-})\right]
^{-1/a}.
\eea

Let us now consider the action of the conformal group $SL(2,R)$, 
realized as the fractional transformation,
\eq\lb{group}
x'^{\pm}= {\a x^{\pm}+\b\over \g x^{\pm}+\d}, \quad \a\d-\b\g=1.
\feq
on the solution (\ref{bhcg}). We find
\eqn\lb{trans}
g_{+-}(x)&=&g'_{+-}(x)\left({{dx'^{+}}\over dx^{+}}\right)^{h}
\left({{dx'^{-}}\over dx^{-}}\right)^{h},\nonumber\\
\e(x)&=&{\e'}(x)\left({{dx'^{+}}\over dx^{+}}\right)^{l}
\left({{dx'^{-}}\over dx^{-}}\right)^{l},
\eea
where $h={1\over 2} (1-{1\over a}), l=-{1\over 2 a}$ are 
``effective'' conformal dimensions of the fields. Notice that we are 
not using the usual definition of conformal dimension of a field, 
which is given by Eqs. (\ref{trans}) with the field on the right hand 
side of the equation evaluated on $x'$ {\sl not} on $x$.
The two definitions agree only for constant fields (e.g for Minkowski
space).  $h$ and $l$ give information about the nontrivial 
$x$-dependence of the conformal fields and characterize their scaling 
behavior.  Our definition is a natural extension to conformal fields
of the geometric notion of isometry (form invariance). For instance, 
AdS$_{2}$ (Eq. (\ref{bh}) with $a=1$) has $SL(2,R)$ as isometry group. 
The usual definition of conformal primary field  keeps track only of 
the tensor character of the metric and assigns to it conformal 
dimension 
equal to 1. Conversely, our definition (\ref{trans}) assigns to 
AdS$_{2}$ zero conformal dimension, consistently with its 
$SL(2,R)$-isometry.
  
For scale transformations ${x'^{\pm}}= \mu x^{\pm}$
($\gamma =\beta =0, \alpha =1/\delta =\sqrt \mu$ in Eq. (\ref{group})) Eqs.
(\ref{trans})
give 
\eq\label{dil}
g_{+-}(x)=\mu^{2h}{g'}_{+-}(x)\quad \e(x)=\mu^{2l}{\e'}(x).
\feq

One can also consider infinitesimal conformal transformations. 
The conformal Killing vectors corresponding to translations, 
dilatations and special conformal transformations are, respectively,
$\chi_{T}^{\pm}=\a,\,\chi_{D}^{\pm}= \mu x^{\pm},\,  
\chi_{S}^{\pm}=-\o  (x^{\pm})^{2}$, whereas Eq. (\ref{trans}) becomes
$\delta g_{+-}=\Omega g_{+-}$, $\delta\e=\Omega \e$, 
with $\O_{T}=0,\,,\O_{D}=2k\mu, \,, \O_{S}=-2k\o(x^{+}+x^{-})$,
$k$ being the above defined conformal dimension of the field.
Only for $a=1$, $a=\infty$ 
(corresponding to the Reissner-Nordstrom case \cite{CCCM}) and $h=0$,
the ground state $M=0$ solution admits full $SL(2,R)$ as isometry group.
For generic values of $a$, the metric admits only the Killing vector
$\chi_{T}^{\pm}$ generating time-translations, but it satisfies 
simple scaling transformations under dilatations and special conformal 
transformations. In the next sections we will relate this conformal 
properties of the vacuum black hole solutions with the scaling 
behavior of the thermodynamical parameters associated with the black 
hole. 
Notice that  also the dilaton $\e$ transforms as a primary field under
$SL(2,R)$. It is (form)-invariant only for $a\to\infty$, whereas it 
transforms as a field of conformal dimension $l$ otherwise.

Eqs. (\ref{bhcg}) hold only for $a\neq 0$.
For $a=0$, introducing appropriate coordinates, the black hole vacuum  
has the form of Minkowski space with a nonconstant dilaton \cite{CM95a}
$ds^{2}=-dy^{+}dy^{-}, \, \e= -(\la^{2}/4)y^{+}y^{-}$.
Obviously, the metric part of this solution behaves under $SL(2,R)$ 
as in Eq. (\ref{trans}) with $h=1$. The dilaton does not follow the 
simple law (\ref{trans})  for all the $SL(2,R)$  transformations.
However under dilatations transforms according to Eq. (\ref{dil})
with $l=1$.

In the following we will need the form of the dilatations in the 
 the coordinate system $(r,t)$ of Eqs. (\ref{bh}). We have, 
\eq\label{scale}
r\to\nu r,\quad t\to\nu^{-a}t, \quad ds^{2}\to\nu^{1-a}ds^{2},
\quad \e\to \nu \e,
\feq
with $\nu=\mu^{-{1 \over a}}$. This formula holds for every value of 
$a>-1$.

The previously described conformal properties of the  $M=0$ solution
do not apply to the $M\neq 0$ black hole solutions. The general 
solution is (time)-translationally invariant but  does not 
satisfy simple scaling laws under the action of 
full $SL(2,R)$.

\section{ Scaling behavior of the thermodynamical parameters}
The physical meaning of the scale symmetry of the black hole vacuum
we have found in the previous section can be easily understood:
the system has no physical scale \footnote{ In the the action (\ref{action}) 
appears the  constant $\la$, which has  the 
dimension of an inverse length. However, $\la$ does not represent a 
dynamical physical scale. It is related with  the size of the curvature 
of the 2D spacetime, but it is not affected by the gravitational 
dynamics}
and we can change the size of the 
system without changing the physical parameters.
The mass $M$, the entropy $S$ and the temperature $T$ of the state are
zero and remain zero if we perform the scale transformations 
(\ref{scale}). 
This is not true anymore when we consider $M\neq 0$ black hole 
solutions.  Now the system has a physical scale (the black hole mass,
or if you prefer the black hole temperature) and the solution does not 
satisfy the simple scale symmetry of Eq. (\ref{scale}) and (\ref{dil}).
We expect that changing the size of the system will affect $M,S,T$.
It is easy to see that the scale symmetry (\ref{scale}) of the black hole 
vacuum 
determines the scaling behavior of the thermodynamical parameters.
The key point is that $M,S,T$ are completely determined by the 
dilaton potential and by the value
$\e_{h}$ of the dilaton at the black hole horizon.
We have 
\eq\label{MST}
M= {\la\over 2}\e_{h}^{a+1}, \qquad S=2\pi\e_{h}, 
\qquad T= {\la\over 4 \pi}(a+1)\e_{h}^{a}.
\feq

Taking into account that the dilaton depends linearly on the coordinate
$r$ (see Eqs. (\ref{bh})), one finds the following scale transformations
\eq\label{scale1}
r\to\nu r,\quad M\to \nu^{a+1}M,\quad S\to \nu S,\quad T\to\nu^{a}T.
\feq
Assuming that $M$ transforms as in Eq. (\ref{scale1}), one finds that 
the scale symmetry (\ref{scale}) is also realized for $M\neq 0$.

The scaling behavior (\ref{scale1}) has  a simple thermodynamical 
interpretation if one considers that for our 2D gravitational system 
$r_{h}$ is the volume $V$.
The scaling  law for the entropy given by  Eq. (\ref{scale1}) 
can be written as $S(\nu V)=\nu S(V)$, meaning that the entropy of the 
black hole is an extensive quantity. For the energy (the black hole 
mass) we have
\eq\label{scaling}
M(\nu V)= \nu^{a+1} M(V).
\feq
Notice that 
we can trade $V$ for $S$ and write $M=M(S)={\la\over 2}({S\over 2\pi})^{a+1}$,
which satisfies $M(\nu S)=\nu^{a+1} M(S)$. 
Hence,  for generic values of $a$ the energy is non-extensive. 
It becomes extensive only for $a=0$, corresponding to the CGHS black 
hole. In this latter case, as it is evident from Eq. (\ref{MST}) the 
temperature is mass-independent and given by $T=\la/4\pi$. The black hole 
geometry  can be described by a Rindler spacetime endowed with a 
nontrivial dilaton and the temperature 
can be expressed in terms of the acceleration of Rindler observers in
Minkowski space.

The non-extensivity of the thermodynamical system defined  by the black 
hole (\ref{bh}) can be also understood as a violation of the Euler 
identity. From Eqs. (\ref{MST}) it follows
\eq\label{euler}
M={1\over a+1} TS.
\feq
The Euler identity is satisfied only for the Rindler case, $a=0$, 
whereas it is violated for generic values of $a$.
It is important to notice that the  non-extensivity of the 
thermodynamical system, expressed  either as the scaling law
(\ref{scaling}) or as the violation of the Euler identity 
(\ref{euler})
can be parametrized in terms of the conformal dimension of the black 
hole vacuum, $h$ (or $l$).

The relations (\ref{scale1}) have been derived using Eqs. 
(\ref{MST}), which express $M,T,S$ as a function of 
the volume of the system. There is however an alternative description 
in which one considers the black hole temperature $T$ as the inverse  
periodicity  at the horizon of the Euclidean time.  In this 
description one continues the black solution (\ref{bh}) into 
Euclidean space, to eliminate the conical singularity at the black 
hole horizon one is forced to put the theory on the cylinder. In this 
way a 
physical scale is generated (the Hawking temperature $T$) and 
its scaling law can be determined using the transformation of the time $t$ 
in Eq. (\ref{scale}) under scale transformations. 

Analytically continuing in Euclidean space the $M\neq 0$ black hole 
solution (\ref{bh}) and introducing an appropriate $R$ coordinate one 
finds near the black hole horizon 
\eq\label{ebh}
ds^{2}= { 4\pi^{2}\over\Theta^{2}}R^{2}dt^{2}+dR^{2},
\feq
where $\Theta=1/T$. In order to eliminate the conical singularity the 
Euclidean time has to be chosen periodic with period $\Theta$.  
The Euclidean description of the black is naturally related with the CFT 
on the cylinder we are going to discuss in the next 
section.  

\section{ Black holes and CFT on the cylinder}

The possibility of describing 2D black holes by means of a CFT has been 
widely investigated in recent years \cite{CM99,ads/cft,cadcav,horizon,liouville}. 
Presently, it is not 
completely clear if it is always possible to mimic the 
gravitational dynamics of the 2D black hole through a CFT. However, in 
some cases and/or for generic black holes in particular regimes, 
CFTs have been shown to give a good description.
This is in particular true for black holes in AdS space, where 
the AdS/CFT correspondence \cite{CM99,ads/cft,cadcav}  should do the job and
for the near-horizon regime of general black holes, for which general 
arguments lead to an associate CFT \cite{horizon}. 

In the 2D dilaton gravity context, CFTs have emerged as an effective 
description of the gravitational system in many cases:  
black holes on AdS$_2$ (the model of Eq. (\ref{bh}) with $a=1$)  
\cite{CM99},
0-branes (the model of Eq. (\ref{bh}) with $a\ge 0$) \cite{CCCM}, Liouville
models \cite{liouville} and there is some evidence that this could also be 
generally true.

In this paper we will assume that our 2D black hole (\ref{bh}) can be 
described by some sort of 2D CFT. We will show that, if this is true, then 
the thermodynamical features of the black hole described in the 
previous section, have a natural interpretation in terms of general
(model independent) features of the CFT.

The description of the black hole given in the previous section can 
be summarized as follows. We have a ground state with simple conformal 
behavior and with no physical scale inscribed, when a black hole of 
mass $M$ is formed a physical scale is generated.  In the Euclidean space,
this physical scale generation is described in geometric terms by considering 
a periodic time, with period $\Theta$. 
It is not difficult to recognize the analogy with   the
relationship between 2D  CFT on the Euclidean plane and CFT on the cylinder,
the mapping between the infinite plane (with holomorphic coordinate 
$z$) and the cylinder (with coordinate $w$ and length of the 
circle $L$) being given by 
\eq \label{mapping}
z=e^{{2\pi\over L}w}.
\feq
It is well-known that if we map the CFT on the plane,
with associated, conformal invariant, vacuum $|0^{pl}>$ and central
charge $c$, into the CFT on the 
cylinder a Casimir energy  $E_{c}$ for the
cylinder vacuum, $|0^{cyl}>$, is generated due to 
finite-size effect (see for instance Ref. \cite{difrancesco}). 
   
Following this analogy we are led to a  
correspondence  between 2D black holes  on the left hand side and 
2D CFT on the right hand side,
\begin{displaymath}
{ black\, hole\, ground\, state \Longleftrightarrow |0^{pl}>}
\end{displaymath}
\begin{displaymath}
M\neq 0\, black\, hole\, state \Longleftrightarrow
|0^{cyl}>
\end{displaymath}
\begin{displaymath}
M\Longleftrightarrow 
E_{c}
\end{displaymath}
\begin{displaymath}
\Theta \Longleftrightarrow L
\end{displaymath}
This  analogy has been already successfully used 
for calculating the entropy of the AdS$_{2}$ black hole in terms of 
degeneracy of states of a 2D CFT \cite{cadcav}. 
The relation between   the $L_{0}$ operator of the Virasoro 
algebra on the plane and on the cylinder is given by (see for 
instance Ref. \cite{difrancesco}),
\eq\label{L0}
L_{0}^{pl}= L_{0}^{cyl}+ {c\over 24}.
\feq
Because we are assuming a conformal invariant vacuum we have
$L_{0}^{pl}|0^{pl}>=0$ and\linebreak 
$L_{0}^{cyl}|0^{cyl}>=0$.
Applying Eq. (\ref{L0}) on the cylinder vacuum $|0^{cyl}>$ we get
\eq\label{vacuum}
L_{0}^{pl}|0^{cyl}>={c\over 24} |0^{cyl}>.
\feq
This means that $|0^{cyl}>$ is an eigenstate of $L_{0}^{pl}$
with eigenvalue $l_{0}^{pl}={c\over 24} $. 

The correspondence between 2D dilaton gravity  and 2D CFT we are 
considering implies that the $|0^{cyl}>$ vacuum has to be considered as 
 an excitation of the $|0^{pl}>$ vacuum. The energy of $|0^{cyl}>$
measured with respect to the $|0^{pl}>$ vacuum is the Casimir energy,
which can be read directly from Eq. (\ref{vacuum}),
$E_{c}={\pi \over 12 L} c$. Identifying the Casimir energy with the 
black hole mass, $E_{c}=M$ and using Eq. (\ref{vacuum})
we can  express  both the central charge and 
$l_{0}^{pl}$ in terms of the black hole mass, 
\eq\label{charge}
c= {12 L\over \pi} M,\qquad  l_{0}^{pl}= {L\over 2\pi} M.
\feq

The Cardy formula \cite{cardy} enables us to calculate the entropy 
associated with the $|0^{cyl}>$ vacuum  considered as an excitation of 
the $|0^{pl}>$ vacuum. Because $|0^{pl}>$ has zero conformal weight
the Cardy formula reads 
\eq\label{cardy}
S=2\pi\sqrt{{c\over 6}l_{0}^{pl}}.
\feq
Feeding Eq. (\ref{cardy}) with Eq.  (\ref{charge}), one finds
\eq
S=2L M,
\feq
which reproduces exactly the black hole thermodynamical relation
(\ref{euler}) after identifying $L$ with $\Theta$ in the following way
\eq
L=\left({a+1\over 2}\right)\Theta.
\feq
This equation implies that the temperature $T$ of the black hole 
cannot 
be identified directly with the inverse of $L$.
A factor, depending on the conformal dimension of the black hole
ground state, relates them. This factor is equal to one only for the
AdS$_{2}$ black hole,  as expected because in this case the 
black hole ground state is truly invariant under $SL(2,R)$.
To end this section let us stress the fact that our derivation of the 
thermodynamical entropy of black holes holds also for asymptotically 
flat black holes. The most striking example is the Schwarzschild black 
hole, which is the particular case $a=-1/2$ of our general model.

\section{Rindler spacetime and CFT}

The analogy between 2D black holes and 2D CFT on the cylinder is 
particularly instructive in two particular cases: $a=1$ corresponding 
to the AdS$_{2}$ black hole and $a=0$ corresponding to the Rindler 
spacetime (CGHS black hole). In both cases the black hole geometry 
is equivalent, modulo space-time diffeomorfisms, to the ground state
$M=0$ solution. The spacetime can be still interpreted as a black 
hole owing to the presence of a nontrivial dilaton \cite{CM95,CM95a}.
The AdS$_{2}$ case has been discussed at length in previous papers
and we will not discuss it here any further.

For $a=0$ the $M=0$ ground state 
solution  can be described, using appropriate coordinates, by 
Minkowski space $ds^{2}= -dy^{+}dy^{-}$. The $M\neq 0$ black hole 
solution has a Rindler form and a $M$-dependent dilaton \cite{CM95a}
\bea\lb{bhcg2}
ds^2 &=& -\exp\left({\la\over 2}(x^{+}-x^{-})\right)
dx^{+}dx^{-}\,,\nonumber\\
\e&=& {2M\over \la}+
\exp\left({\la\over 2}(x^{+}-x^{-})\right).
\eea
The coordinates $x^{\pm}$ give a Rindler coordinatization of Minkowski 
space,
\eq
y^{\pm}= \pm{2\over \la}\exp\left(\pm {\la\over 2} x^{\pm}\right).
\feq
It is not difficult to recognize in the previous equation the 
Minkowskian version of the mapping between CFT on the plane and CFT on 
the cylinder given in Eq. (\ref{mapping}).
In the Rindler case the black hole/CFT correspondence  presented in the 
previous section is more stringent then in the general case. The 
transformation that maps  vacua in the CFT is the same as the
map between  the  ground state and the black hole.
It is clear that also for  $a$  generic one can show that the 
mapping (\ref{mapping}) maps the Euclidean black hole metric (\ref{ebh})
into the Euclidean plane
 (which is conformally related to the black hole ground state).
This is a simple consequence  of  the fact that Eq. (\ref{ebh}) is a
near-horizon expansion. Near the horizon the generic black hole 
solution (\ref{bh}) has always a Rindler form. But, this  is true
only locally, near the horizon, whereas in the pure Rindler case the 
mapping holds globally, for the whole space. 

The Minkowski vacuum $|0^{M}>$ of  the gravitational description
corresponds to the $|0^{pl}>$ vacuum in the CFT description, whereas 
the Rindler vacuum $|0^{R}>$ corresponds to $|0^{cyl}>$.
The well-known fact that the Rindler observer  will see $|0^{M}>$ as filled with 
thermal radiation with temperature $T_{R}=\la/4\pi$ can be interpreted from  
the CFT point of view as a finite-size effect, with the length $L$ 
related to the Rindler temperature by $L= 1/2 T_{R}$.
Moreover our correspondence can be used to assign in a natural way 
mass and entropy to the Rindler spacetime. The Rindler mass  and 
entropy can be expressed in terms of   
the central charge of the CFT. The mass is simply given by 
the Casimir energy of $|0^{cyl}>$, $M_{R}=E_{c}$, whereas the entropy 
is  $S_{R}=M_{R}/ T_{R}= (4\pi/\la)E_{c}$.

\end{document}